\documentclass[12pt]{iopart}
\usepackage{graphicx}
\usepackage{epsfig}
\usepackage{here}
\usepackage{dcolumn}
\usepackage{bm}
\begin{document}
\title{Nonlinearity-assisted quantum tunneling in a matter-wave interferometer}

\author{Chaohong Lee, Elena A. Ostrovskaya and Yuri S. Kivshar}

\address{Nonlinear Physics Centre and ARC Centre of Excellence for Quantum-Atom
Optics, Research School of Physical Sciences and Engineering,
Australian National University, Canberra ACT 0200, Australia}

\begin{abstract}
We investigate the {\em nonlinearity-assisted quantum tunneling}
and formation of nonlinear collective excitations in a matter-wave interferometer, which is realised by the
adiabatic transformation of a double-well potential into a
single-well harmonic trap. In contrast to the linear quantum tunneling
induced by the crossing (or avoided crossing) of neighbouring
energy levels, the quantum tunneling between different nonlinear
eigenstates is assisted by the nonlinear mean-field interaction.
When the barrier between the wells decreases, the mean-field
interaction aids quantum tunneling between the ground and excited
nonlinear eigenstates. The resulting {\em non-adiabatic evolution}
depends on the input states. The tunneling process leads to the
generation of dark solitons, and the number of the generated dark
solitons is highly sensitive to the matter-wave nonlinearity. The
results of the numerical simulations of the matter-wave dynamics
are successfully interpreted with a coupled-mode theory for
multiple nonlinear eigenstates.
\end{abstract}

 \submitto{\jpb} \maketitle

\section{Introduction}

Matter-wave interferometry involves coherent manipulation of
the external or internal degrees of freedom of massive particles
\cite{Berman-book,Nature-Steven-Chu}. Utilizing the well-developed
techniques of trapping and cooling, the matter-wave
interferometers have been realised with atomic Bose-Einstein
condensates (BECs)
\cite{MIT-doublewell-BEC,Heidelberg-doublewell-BEC,Austria-doublewell-BEC,SUT-doublewell-BEC}.
Almost all BEC interferometers based on spatial interference
measure the phase coherence by merging two initially separated
condensates
\cite{MIT-doublewell-BEC,Heidelberg-doublewell-BEC,Austria-doublewell-BEC,SUT-doublewell-BEC}.
To recombine two condensates confined in a double-well potential,
one has to transform the double-well potential into a single-well
potential by decreasing the barrier height.

Intrinsic interparticle interactions in atomic condensates have
stimulated various studies of the nonlinear behaviour of condensed
atoms~\cite{Nature-Phillips}. A balance between matter-wave
dispersion and nonlinear interaction supports a number of
nontrivial collective excitations, including bright solitons in
condensates with attractive interactions
\cite{Science-bright-soliton,Nature-bright-soliton} and dark
solitons in condensates with repulsive interactions
\cite{PRL-dark-soliton,Science-dark-soliton}.  In a harmonically trapped condensate with repulsive interparticle
interactions, the nodes of excited nonlinear eigenstates
correspond to dark solitons~\cite{Tristram}, so that the formation
of dark solitons can be associated with populating excited
states~\cite{Tristram,yukalov}. Several methods of condensate
excitation have been suggested, the most experimentally appealing
ones involving time-dependent modifications of trapping potentials
\cite{damski}. The operation of BEC interferometers and splitters
based on spatiotemporal Y- and X-junctions
\cite{PRA-Zozulya,PRL-MIT-phase-sensitive-recombination} is
greatly affected by the possibility of nonlinear excitations. The nonlinear excitations in BEC
interferometers with repulsive interparticle interactions lead to
the generation of dark solitons
\cite{Reinhardt-JPB,PRL-MIT-phase-sensitive-recombination}, and
can be utilised to enhance the phase sensitivity of the devices
\cite{PRL-MIT-phase-sensitive-recombination,negretti}.

The extensively explored mechanisms for population transfer between different eigenstates
of a trapped BEC include non-adiabatic
processes \cite{PRA-Zozulya}, Josephson tunneling
\cite{PRL-Smerzi,PRA-Bose-Josephson-Lee}, and Landau-Zener
tunneling
\cite{LZ-tunneling-Liu1,LZ-tunneling-Liu2,LZ-tunneling-Korsch},
which are also responsible for population transfer in linear
systems. However, in a sharp contrast to linear systems, the
quantum tunneling between different nonlinear eigenstates can be
assisted by the nonlinear mean-field interaction even in the
absence of crossing (and avoided crossing) of the energy levels.
Up to now, this peculiar type of quantum tunneling remains poorly
explored.

In this paper, we explore the intrinsic mechanism for the quantum
tunneling assisted by repulsive nonlinear mean-field interactions
in a matter-wave interferometer. We consider the dynamical
recombination process of a BEC interferometer, in which an
initially deep one-dimensional (1D) double-well potential is
slowly transformed into a single-well harmonic trap. Our numerical
simulations, employing a time-dependent 1D mean-field
Gross-Pitaevskii (GP) equation, show that multiple moving dark
solitons are generated as a result of the nonlinearity-assisted
quantum tunneling between the ground and excited nonlinear
eigenstates of the system, and the qualitative mechanism is
independent on the particular shape of the symmetric double-well
potential. Furthermore, the number of the generated dark solitons
is found to be highly sensitive to the strength of the effective
nonlinearity that in turn depends on the total number of condensed
atoms and the atom-atom s-wave scattering length. The population
transfer between different nonlinear eigenstates caused by the
nonlinearity-assisted quantum tunneling can be quantified by a
coupled-mode theory for multiple nonlinear eigenstates of the
system.

\begin{figure}[ht]
\centerline{\includegraphics[width=18cm,height=8cm,keepaspectratio]
{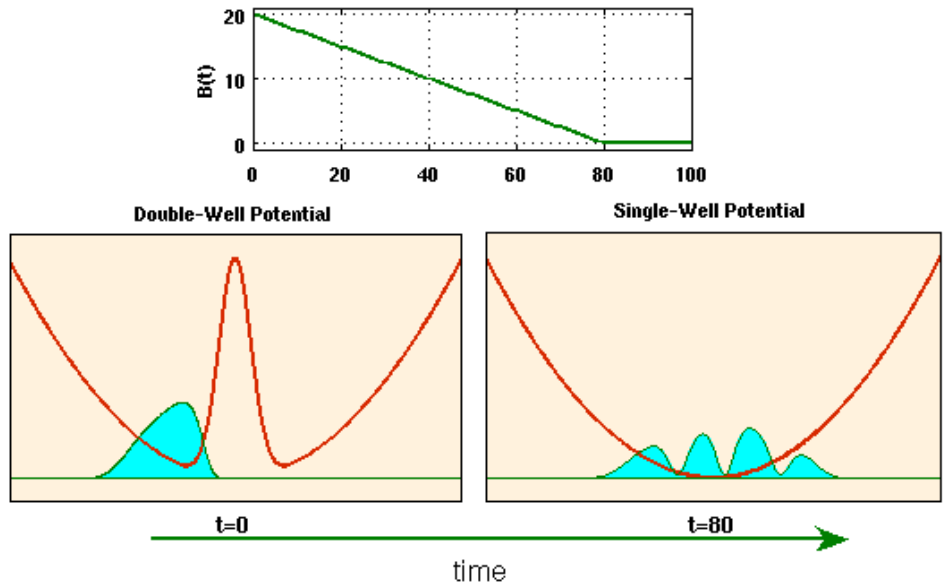}} \caption{Schematic diagram. Top: time-dependence of
the barrier height $B(t)$. Bottom left: initial BEC density
distribution (shaded) at $t = 0$ in the double-well potential
(solid line). Bottom right: density distribution (shaded) at $t =
80$ in the single-well potential (solid line).} \label{fig.1}
\end{figure}

\section{Model and numerical results}

We consider a condensate under strong transverse confinement, $m
\omega^{2}_{\rho} (y^{2}+z^{2})/2$, so that the 3D
mean-field model can be reduced to the following 1D model
\cite{1d_reduction}:
\begin{equation}
i \hbar \frac{\partial}{\partial t} \Psi (x,t) = H_{0} \Psi (x,t)
+ \lambda \left|\Psi (x,t) \right|^{2} \Psi (x,t),
\end{equation}
where $H_0 = -(\hbar^{2}/2m)(\partial^{2}/\partial x^{2}) +
V(x,t)$, $m$ is the atomic mass, $\lambda>0$ characterizes the
effective nonlinearity which we assume to be {\em repulsive}, and
$V(x,t)$ is an external potential. If the condensate order
parameter $\Psi(x,t)$ is normalised to one, the effective
nonlinearity $\lambda = 2N a_{s} \omega_{\rho} \hbar$ is
determined by the total number of atoms $N$, the s-wave scattering
length $a_{s}$ and the transverse trapping frequency
$\omega_{\rho}$ \cite{PRL-Carr-Brand-soliton}. In what follows we
use the dimensionless version of the model equation obtained by
choosing the natural units of $m = \hbar =1$.

\begin{figure}[ht]
\centerline{\includegraphics[width=18cm,height=10cm,keepaspectratio]{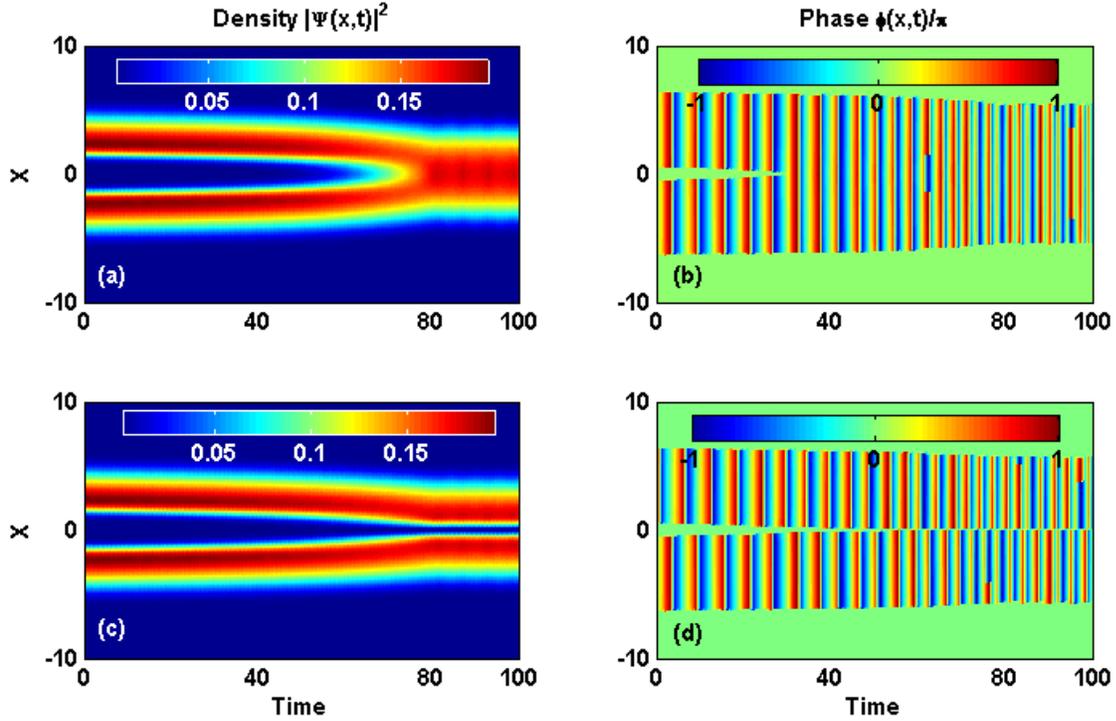}}
\caption{Evolution of the condensate density (left) and phase
(right) of the ground and first-excited states for the system of
the effective nonlinearity $\lambda = 20$. The first and second
rows correspond to the evolution of the ground and first-excited
states, respectively.}\label{fig.2}
\end{figure}

We assume the time-dependent potential $V(x,t)$ as a
spatiotemporal Y-shape potential generated by the superposition of
a 1D time-independent harmonic potential and a time-dependent
Gaussian barrier (see Fig. 1):
\begin{equation}
 V(x,t) = \frac{1}{2} \omega^{2} x^{2} + B(t) \cdot \exp \left(-\frac{x^{2}}{2d^{2}}\right),
\end{equation}
where $\omega$ is the trapping frequency, $d$ is the barrier width, and
the barrier height depends on time as follows:
\begin{equation}
B(t)= \left\{
\begin{array}{ll}
B_{0} - \alpha t, &\textrm{for~~~} t < B_{0} / \alpha, \\
0, &\textrm{for~~~} t \geq B_{0} / \alpha,
\end{array}
\right.
\end{equation}
where $\alpha$ is the rate at which the barrier between the wells is ramped down.
 When the barrier height
$B(t)>\omega^2d^2$, the time-dependent potential is a double-well
potential with two minima at $x=\pm d
\sqrt{2\ln[B_0/(d^2\omega^2)]}$. Thus, the 1D description is valid
for weak longitudinal confinements satisfying
$\omega^{\textmd{db}} = \omega \sqrt{2\ln[B_0/(d^2\omega^2)]} \ll
\omega_{\rho}$ and $\omega \ll \omega_{\rho}$. To ensure the
adiabatic evolutions of the symmetric and antisymmetric initial
eigenstates, the rate $\alpha$ must be sufficiently
small. In Fig. 2, we show the evolutions of the ground and
first-excited eigenstates of the system of the effective nonlinearity
$\lambda = 20$ and ramping rate $\alpha = 1/4$. For such a small
value of $\alpha$, both the ground and first excited eigenstates of the
initial double-well potential adiabatically evolve into the
corresponding ground and first excited eigenstates of the final
single-well potential. This means that the {\em non-adiabatic
effects are negligible} for such a slowly varying process.

The usual double-well BEC interferometers involve condensates
trapped in a double-well potential before recombination. Below, we
consider the case of the initial state of the BEC being fully
localised in a single well of a symmetric double-well potential
with a sufficiently high barrier, so that there is {\em no
significant overlap} between the Wannier states of the two wells.
The fully localised initial state can be viewed as the
equal-probability superposition of the ground and first-excited
eigenstates, so that it can be used to observe the interference of
these two eigenstates \cite{RZ-tunneling-Fu}. For BECs trapped in
such a deep potential, the mean-field ground and first excited
states are degenerate or quasi-degenerate. Even for low barriers,
if the tight-binding condition is still satisfied, the two-mode
approximation will give the picture of a classical Bose-Josephson
junction. In the framework of second quantisation, the system
obeys a two-site Bose-Hubbard Hamiltonian. In this fully quantum
picture, a completely localised initial state corresponds to the
highest excited state for repulsive interactions. This state exhibits
degeneracy with the sub-highest excited
state~\cite{PRL-Lee-AMZIQBJJ}, which corresponds to the
bistability in a classical Bose-Josephson junction~\cite{PRA-Lee}.

\begin{figure}[ht]
\centerline{\includegraphics[width=18cm,height=12cm,keepaspectratio]{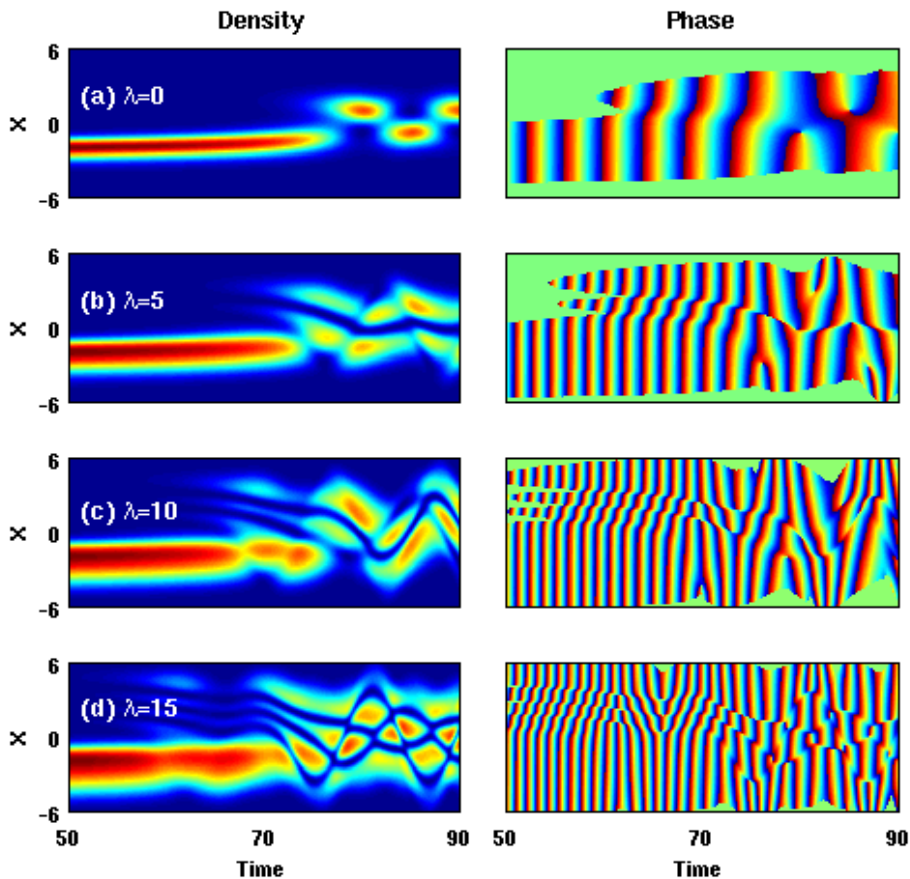}}
\caption{Evolution of the condensate density (left) and phase
(right) for different values of the effective nonlinearity
$\lambda$. Cases (a-d) correspond to $\lambda = 0$, $5$, $10$ and
$15$, respectively.}\label{fig.3}
\end{figure}

Since initially there is no overlap between the two Wannier
states, the quantum tunneling between those states is negligible.
As the barrier height gradually decreases, the overlap between two
Wannier states becomes more significant. Then both the
quasi-degeneracy between the ground and first excited states in
the mean-field picture and the quasi-degeneracy between the
highest excited and sub-highest excited states in the quantum
picture break down. The quantum tunneling of the fully localised
state becomes more pronounced as the barrier height is decreasing.
Due to the very slow reduction in the barrier height, the kinetic
energy stays small during the whole process, and the quantum
tunnelling dominates the dynamics. However, the over-barrier
hopping could occur in a rapidly varying process, in which case the kinetic
energy can exceed the potential barrier.

To explore the dynamic
evolution, we numerically integrate the GP equation with the
well-developed operator-splitting procedure and the absorbing
boundary conditions. In Fig. 3 we show the time evolution of the condensate density
and phase for the trapping frequency $\omega = 0.2 \pi$, the
initial barrier height $B_{0} = 20.0$, the barrier width $d =
\sqrt{2}/2$, the ramping rate $\alpha = 1/4$, and different values
of the effective nonlinearity $\lambda$.  For the chosen small ramping down rate, all
symmetric (antisymmetric) states of the deep double-well
potential will adiabatically evolve into the corresponding ground
(or excited) states of the single-well potential.

\begin{figure}[ht]
\centerline{\includegraphics[width=18cm,height=10cm,keepaspectratio]{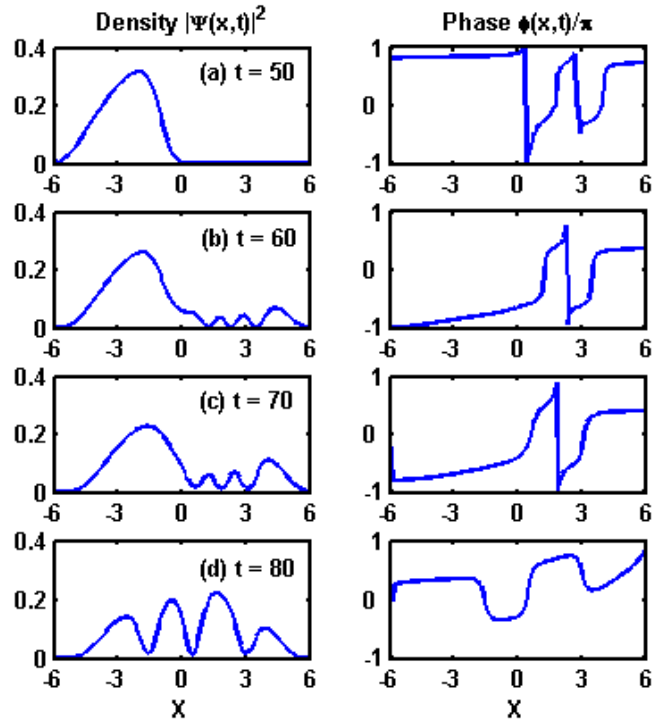}}
\caption{Formation of dark solitons in the system with the
effective nonlinearity $\lambda = 15$. Left: density distributions
$|\Psi(x,t)|^{2}$ for different times. Right: phase distributions
$\phi(x,t)$ for the corresponding density distributions.}
\label{fig.4}
\end{figure}

Evolution of the fully localised initial state strongly depends on
the values of the effective nonlinearity $\lambda$. In the linear
case ($\lambda = 0$), the fully localised initial state can be
viewed as an equal-probability superposition of the symmetric and
antisymmetric states, so that, according to the adiabatic theorem, the evolving state is always the
equal-probability superposition of the ground and the first
excited states of the system. Due to the nonlinear interactions, the superposition principle
becomes invalid, and the resulting behavior can be interpreted as
the coupled dynamics of the ground and multiple excited states of
the nonlinear system. Akin to the linear systems, the quantum
tunneling appears once the quasi-degeneracy between the ground and
the first excited state is broken, and gradually becomes
significant with decreasing barrier height. The time scale on
which the quantum tunneling appears in the nonlinear system
shortens with the growth of nonlinear interaction strength $\lambda$. The
excited nonlinear states of the BEC  in a single-well potential
can be thought as stationary configurations of single or multiple
dark solitons \cite{Tristram}. As a result of the population
transfer to such excited modes, the condensate develops multi-peak
distribution with significant phase gradients across density notches
between neighboring peaks. These notches are dark or gray solitons
with well-defined phase gradients close to $\pi$ (see Fig. 4).

The number of dark solitons formed in this process varies with the effective nonlinearity $\lambda$ (see Fig. 5). This dependence exhibits multiple plateaus as the
effective nonlinearity $\lambda$ changes, as shown in Fig. 5. Given the
relationship between the nonlinear interaction strength and the key
parameters of the system, $\lambda = 2N a_{s} \omega_{\rho}
\hbar$, one can control the number of generated solitons by
adjusting the s-wave scattering length with the Feshbach
resonance, the total number of atoms in the condensate with
initial preparation, and/or the transverse trapping frequency by
tuning the transverse trapping field strength. The number of
solitons remains unchanged for a long period of time before multiple
collisions between solitons take place. The inelastic collisions lead to the
radiation of small-amplitude waves, and after a large number of collisions the number of solitons oscillating in the trap changes.

\begin{figure}[ht]
\centerline{\includegraphics[width=18cm,height=6cm,keepaspectratio]
{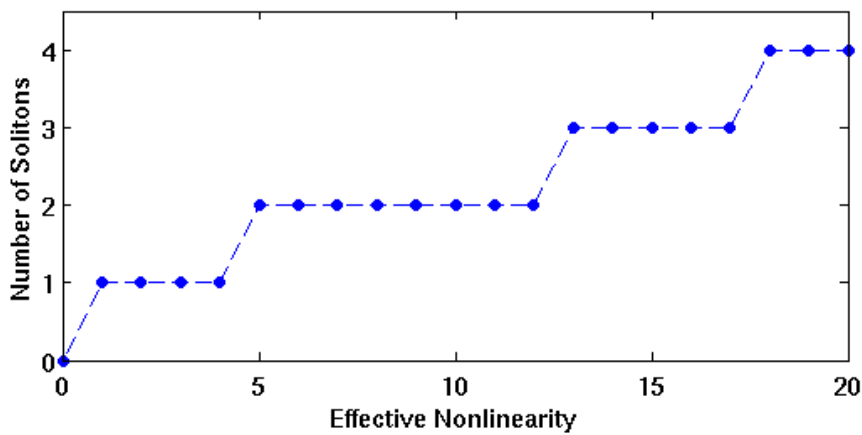}} \caption{Number of dark solitons generated in the
condensate at $t = 80$ versus the effective nonlinearity parameter
$\lambda$.} \label{fig.5}
\end{figure}

\section{Modal decomposition}

To obtain the quantitative picture of the population transfer, we
decompose an arbitrary state of our time-dependent system as
\cite{negretti,PRA-elena},
\begin{equation}
\Psi(x,t) = \sum _{j}^{N} C_{j}(t) \phi_{j}(x,t),
\end{equation}
where $\phi_{j}(x,t)$ is the $j$-th stationary state for the
nonlinear system with the potential $V(x,t)$, which obeys the
(dimensionless) equation:
\begin{equation}
\mu_{j}(t) \phi_{j}(x,t) = \left[ -\frac{1}{2}
\frac{d^{2}}{dx^{2}} + V(x,t)\right] \phi_{j}(x,t) + \lambda
\phi^{3}_{j}(x,t).
\end{equation}
Here $\mu_{j}(t)$ is the chemical potential for the $j$-th
stationary state. For each instantaneous form of the time-dependent potential, the nonlinear eigenstates
$\{\phi_{j}(x,t)\}$ form an orthogonal set,  similarly to their linear counterparts \cite{negretti,Tristram, PRA-Elena}. Due to the nonlinear interparticle interactions, there exist
additional stationary states (e.g.,  self-trapped states) which
have no linear counterparts \cite{PRA-Elena,dagosta}. Nevertheless, every stationary state of the nonlinear system can be composed from the orthogonal basis
$\{\phi_{j}(x,t)\}$. The nonlinear eigenstates of
time-independent potentials are also time-independent. However, due
to the violation of the superposition principle, population transfer
between different nonlinear eigenstates also occurs in time-independent
systems \cite{PRA-Elena}. This type of population transfer, which
originates from the exchange collisions between atoms in different
eigenstates, will be considered in detail below.

The population dynamics for different nonlinear eigenstates can be
described by the evolution of the complex coefficients $C_{j}(t)$,
which obey a series of coupled first-order differential equations,
\begin{equation}
\label{cm_eq}
 i\frac{dC_{l}(t)}{dt} = \sum_{j}^{N} \left[
E_{0}^{l,j} + \sum_{k,k^{\prime}} Q_{k,k^{\prime}}^{l,j}
C_{k}^{*}(t)C_{k^{\prime}}(t) \right]C_{j}(t).
\end{equation}
Due to the conservation of the total number of particles,
$C_{j}(t)$ satisfy the normalisation condition $\sum_{j}
|C_{j}(t)|^2 = 1$. Here the linear coupling parameters are
\begin{equation}
E_{0}^{l,j}(t) = \int \phi^{*}_{l}(x,t) H_{0} \phi_{j}(x,t) dx,
\end{equation}
and the nonlinear coupling parameters are
\begin{equation}
Q_{k,k^{\prime}}^{l,j}(t) = \lambda \int \phi^{*}_{l}(x,t)
\phi^{*}_{k}(x,t) \phi_{k^{\prime}}(x,t) \phi_{j}(x,t) dx.
\end{equation}
For a spatially symmetric potential $V(x,t)=V(-x,t)$, we have
$Q_{k,k^{\prime}}^{l,j}(t)=0$, and then $(k + k^{\prime} + l + j)$
are odd integer numbers.

In our numerical simulations, we generalise the direct relaxation
method for linear quantum systems \cite{Kosloff} to calculate the
eigenstates and their eigenvalues (chemical potentials) for our
nonlinear system with different effective nonlinearities at any
moment of time. Projecting the condensate wavefunction $\Psi
(x,t)$ onto the nonlinear eigenstates $\phi_{j}(x,t)$, we find the
population probabilities $P_{j}(t) = |C_{j}(t)|^{2} = |\int
\phi_{j}^{*}(x,t)\Psi (x,t)dx|^{2}$ which depend on the time and
the effective nonlinearity $\lambda$.

\begin{figure}[ht]
\centerline{\includegraphics[width=18cm,height=10cm,keepaspectratio]{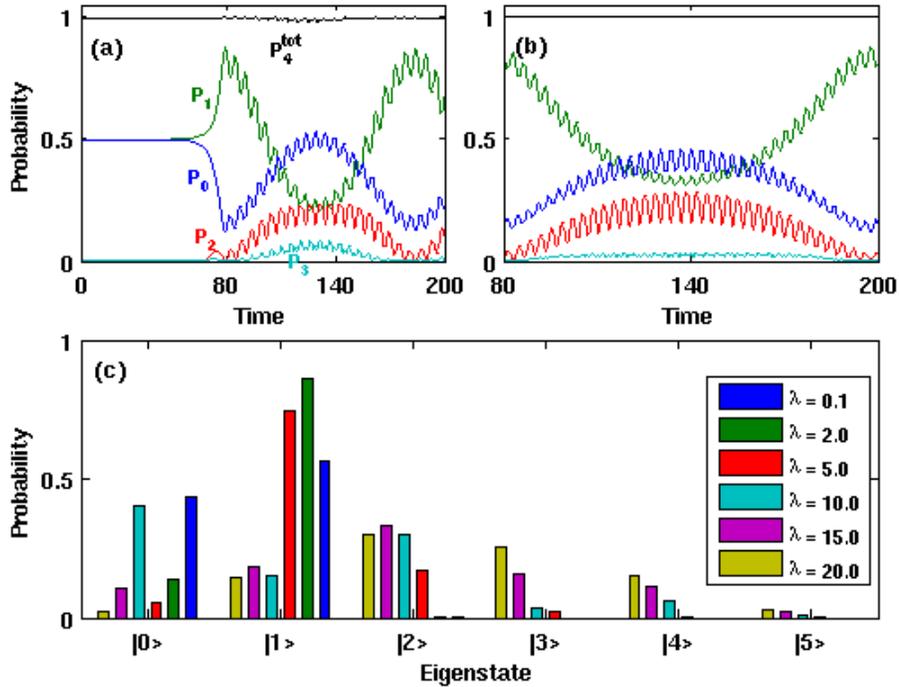}}
\caption{(a) Time evolution of population probabilities in
different eigenstates for $\lambda = 2$,  where $P_0$ and $P_j$
$(j=1, 2, 3)$ are the population probabilities of the ground state
and the $j-$th excited state, respectively. The  total probability
of the first four eigenstates is denoted as $P_{tot} = P_0 + P_1 +
P_2 + P_3$. (b) The corresponding population evolution for $t>80$
obtained from the coupled-mode equation (6) with first four lowest
eigenstates. (c) Population probabilities for the first six lowest
nonlinear eigenstates at $t=80$.} \label{fig.6}
\end{figure}

In Fig. 6(a), we show the time evolution of the population
probabilities $P_{j}(t)$ for the effective nonlinearity $\lambda =
2.0$. Here we only consider four lowest eigenstates (i.e., $N=4$),
so that $P_0$ is the ground state population probability, $P_j$
$(j=1, 2, 3)$ are the population probabilities of the $j-$th
excited state, and $P_{\rm tot} = P_0 + P_1 + P_2 + P_3$ is the
total probability of the first four eigenstates. For $t < 50$, the
population probabilities keep almost unchanged. In the region of
$50 < t < 80$, we observe a fast population transfer from the
ground state to the first excited state. After the recombination
of the two wells, i.e. for $t>80$, the populations in different
nonlinear eigenstates oscillate with time, even though the nonlinear system has
a time-independent potential, time-independent eigenstates, and no
energy degeneracy between neighbouring eigenstates. This behavior differs
drastically from the linear dynamics where populations in
different eigenstates always remain unchanged. We find that at this
value of the effective nonlinearity the total population
probability $P_{\rm tot}(t)$ in the first four eigenstates is
always close to one. The low-frequency population oscillations are
dominated by the linear coupling between different modes and the
high-frequency ones are due to the nonlinear cross-coupling of the
nonlinear modes which corresponds to the exchange collision of
atoms in different eigenstates.

For small $\lambda$, the dynamics of $P_{j}$ after the merging of
the two wells can be approximately captured by the projection of
the BEC state at the moment of the merging (here $t=80$) onto the
set of $N$ stationary nonlinear states $\phi_j(x)$ of the
single-well potential $V_0(x)$ of $B(t)=0$. This is confirmed in
Fig. 6(b), where we employ the coupled-mode theory (\ref{cm_eq})
with $N=4$ eigenstates of $V_0(x)$ [cf. Fig. 6(a)]. The number of
eigenstates, $N$, that must be considered in the coupled-mode
theory, increases with the effective nonlinearity. The
highest-order mode ($j=N$) of the harmonic potential $V_0(x)$ with
significant (non-zero) excitation probability $P_N$ at the merging
time will therefore determine the number $N$ of dark solitons that
are likely to be formed. Figure 6(c) shows the excitation
probabilities at $t=80$ for the first six lowest eigenstates of
$V_0(x)$ and different $\lambda$. By comparing the number of
significantly excited states for different values of $\lambda$,
one can see that the number of dark solitons formed is indeed
approximately determined by the highest-excited nonlinear mode of
the harmonic trap that is still sufficiently populated. For
instance, $N=1$ solitons are expected to form for $\lambda=2$, and
$N=2$ for $\lambda=10$ (cf. Fig. 5).

\section{Conclusions}

We have explored the nonlinearity-assisted quantum tunneling and formation of nonlinear collective excitations in the matter-wave interferometer based on a time-dependent double-well potential, dynamically reconfigured to form a single-well harmonic trap . In
contrast to the Josephson tunneling and Landau-Zener tunneling,
the nonlinearity-assisted quantum tunneling is brought about by
the nonlinear inter-mode population exchange scattering. The
excitations caused by this type of tunneling lead to the dark
soliton generation in the process that differs dramatically from
the phase imprinting \cite{PRL-dark-soliton,Science-dark-soliton}
or condensates collisions \cite{Reinhardt-JPB}. The number of
generated solitons can serve as a sensitive measure of the degree
of the nonlinearity in the system. With the well-developed
techniques for preparing and manipulating condensed atoms in
double-well potentials
\cite{MIT-doublewell-BEC,Heidelberg-doublewell-BEC,Austria-doublewell-BEC,SUT-doublewell-BEC},
loading the condensed atoms in one well of a deep double-well
potential and adjusting the barrier height, the experimental
observation of this effect seems feasible.

{\it Note added:} After this manuscript was prepared for
submission, the group of Peter Engels from the Washington State
University reported the experimental observation of matter-wave
dark solitons due to quantum tunneling \cite{Engels-experiment},
in the process of sweeping a high potential barrier from one edge
of the trap to the other.

The authors thank R. Gati and M. Oberthaler for stimulating
discussions. This work was supported by the Australian Research
Council (ARC).

\end{document}